# Near-zero-index waveguide for beam steering

**Chih-Zong Deng,*** **Eri Igarashi, and Yoshihiro Honda**
Sony Group Corporation, Technology Infrastructure Center, Advanced Research Laboratory, 4-14-1 Asahi-cho, Atsugi-shi, Japan, 243-0014

**Abstract**. Zero-index materials (ZIMs) have become popular because of their unique optical behaviors, such as infinite effective wavelengths and spatially uniform electromagnetic distributions. The all-dielectric ZIMs—Dirac-like cone-based zero-index materials (DCZIMs)—are used in various photonic applications owing to their superior optical properties, i.e., finite impedances, zero Ohmic losses, and high compatibility with photonic circuits. We propose a more general and simple approach that is independent of the Dirac-like cone mode for realizing near-zero index (NZI) behavior in all-dielectric waveguides. This approach can be applied to various dielectric materials up to the necessary NZI bandwidth. $Si_3N_4$ and Ge NZI waveguides are demonstrated for achieving broadband and narrowband NZI, respectively. The proposed broadband NZI waveguide achieves a bandwidth of 140 nm for $|n_{eff}| < 0.1$ ($n_{eff}$ = effective refractive index) at telecommunication wavelengths, which is 2 times larger than that of the reported NZI waveguides. Further, NZI waveguide-based beam steering was demonstrated with a wide steering range ($\Delta\theta \approx 105°$) across the radiation angle of $\theta = 0°$. The proposed NZI-waveguide design principle and beam steering present a feasible approach for the development of photonic circuits and zero-index-based photonic applications.

## 1 Introduction

Metamaterials, which are artificial materials composed of subwavelength-scale unit cells, are at the forefront of nanophotonics research because of their unique optical properties that are not observed in natural materials.[1] Photonic metamaterials have emerged as promising solutions for overcoming the challenges of photonic devices, such as miniaturization[2] and weak light-matter interactions[3]. A Dirac-like cone-based zero-index material (DCZIM) is a type of metamaterial that realizes zero-index properties by simultaneously achieving zero magnetic and electric responses at a given frequency; these zero responses are realized by adjusting the dielectric resonators. DCZIMs have attracted considerable attention owing to their unique properties, such as infinite effective wavelengths and uniform spatial phase distributions over the entire medium.[3-15] Compared to the widely studied ZIMs, i.e., the epsilon-near-zero ZIMs, the DCZIMs exhibit superior optical properties, including finite impedances and zero Ohmic losses, owing to both zero permittivity (ε) and permeability (μ) at a given frequency and their all-dielectric structure. DCZIMs have been used in various photonic applications, such as cloaking[4], displacement measurements,[5] lensing,[6] and beam steering.[7] Although 2D-DCZIMs (photonic crystals) have been extensively studied in recent years,[3-13] only a limited number of reports[14-15] are available on 1D-DCZIMs (waveguides).

Recently, near-zero-index (NZI) waveguides[14-15] based on the Dirac-like cone mode have been demonstrated. However, the design principle of the current NZI waveguides is based on 2D metamaterials with Dirac-like cone dispersions at the center of the Brillouin zone, which requires accurate band-structure calculations with extremely low tolerances[8,12]. Here, we propose a new, more general, and simple design principle, not based on complicated band-structure calculations, for realizing NZI waveguides. The beam-steering function can be realized by utilizing the wavelength-dependent radiation of the NZI waveguide. The waveguides exhibit broadband (140 nm, for $|n_{eff}| < 0.1$; here, $n_{eff}$ is the effective refractive index) or narrowband NZI (50 nm, for $|n_{eff}| < 0.1$) at telecommunication wavelengths by exploiting the low-refractive-index material $Si_3N_4$ or high-refractive-index material Ge, respectively. Notably, the bandwidth of the broadband NZI is more than twice of that reported in previous studies.[10,14] Taking advantage of the continuous angular dispersion across the zero-index mode, radiation in the forward to backward direction can be achieved by simply changing the wavelength of the incident light.

## 2 Design principle of near-zero-index waveguide

Let us consider a light source that emits light in a ZIM that is surrounded by an arbitrary medium with a refractive $n_2$. According to Snell's law,

$$n_1 \sin \theta_1 = n_2 \sin \theta_2 \quad (1)$$

The light will always leave the surface of the ZIM perpendicularly ($\theta_2 = 0$) irrespective of the incident angle $\theta_1$, because of $n_1 = 0$, as shown in Fig. 1(a). To realize an NZI waveguide, a periodic air-hole structure was applied on the waveguide, as shown in Fig. 1(b). Fig. 1(c) illustrates the wave behavior when light is incident from the bottom of the NZI waveguide, along the *y*-axis, at the resonant wavelength. Because the propagation mode in the NZI waveguide is a radiative mode that operates above the light line, the light couples to free space while propagating in the waveguide. Guide wavelength ($\lambda_{guide}$) is defined as the distance between two equal phases along the waveguide. When the guide wavelength $\lambda_{guide}$ is equal to the period of the period air-hole structure, the propagating light couples to the free-space light through the air-holes with the same phase throughout the NZI waveguide, as illustrated in Fig. 1(c) and (i). Thus, for a properly optimized NZI waveguide, when light propagates in the NZI waveguide at a resonant wavelength, it will always couple with the free-space light through the air-hole with the same phase. As a result, continuous plane-wave radiation will be perpendicularly emitted from the NZI waveguide, as illustrated in Fig. 1(c) and (iii). To obtain the guide wavelength $\lambda_{guide}$, light was launched from the bottom of the NZI waveguide along the y-axis with different wavelengths $\lambda_0$ in a finite-difference time-domain simulation (FullWAVE, RSoft Design Group). The NZI waveguide was designed as a periodic air-hole structure on a Si waveguide with a period of 660 nm, radius of 150 nm, and thickness of 220 nm. The calculated guide wavelength $\lambda_{guide}$ is shown in Fig. 1(d). When $\lambda_0 = 1550$ nm, $\lambda_{guide} = 660$ nm, which is the same as the period of the periodic air-hole structure. As mentioned above, the waveguide exhibits zero-index properties when the period $= \lambda_{guide}$. Therefore, $\lambda_0 = 1550$ nm was determined to be the resonant wavelength for realizing a zero index for the designed NZI waveguide. To validate the zero-index behavior, light was launched into the NZI waveguide and propagated along the *y*-direction with $\lambda_0 = 1550$ nm. Fig. 1(e) shows the distribution of the real part (Re) of the magnetic field $H_z$ in the $xy$-plane in the middle of the NZI waveguide. The light was coupled to the free-space light as plane waves that radiated perpendicular to the NZI waveguide. The phases of the propagating light were the same for every air-hole position, which enabled the formation of perpendicular radiation, as clearly seen in Fig. 1(e) and (ii). Perpendicular radiation indicates the zero-index property of the designed waveguide.

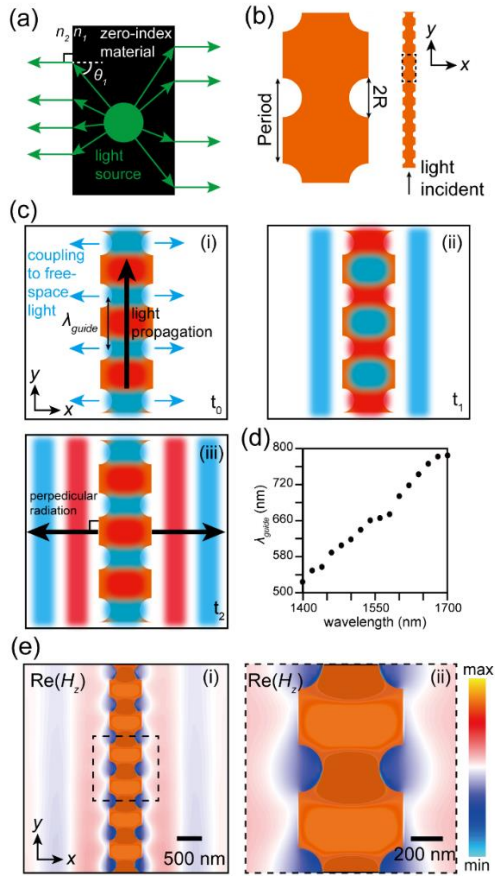

**Fig. 1.** (a) Schematic of a light source that emits light in a ZIM. (b) Schematic of an NZI waveguide. (c) Schematic of the light propagating in the NZI waveguide and coupling to the free-space light at resonant wavelength. (d) Calculated guide wavelength $\lambda_{guide}$ as a function of the incident light wavelength $\lambda_0$. (e) Distribution of the real part of the magnetic field $H_z$ in the $xy$-plane at the middle of the NZI waveguide.

## 3   Characterization of near-zero-index waveguide

To characterize the proposed NZI waveguide, incident light of different wavelengths was used in the simulation. The real part and phase distribution of the magnetic field $H_z$ at different wavelengths are shown in Fig. 2(a). Phase-advance free propagation was observed at $\lambda_0 = 1550$ nm, indicating the zero-index wave behavior. Further, the phase-advance and phase-delay modes, which originate from the mismatch between $\lambda_{guide}$ and the period of the structure, were observed at $\lambda_0 = 1450$ and 1650 nm, respectively. When $\lambda_0 = 1450$ nm, $\lambda_{guide}$ is 575 nm, which is shorter than the period; thus, the phase-advance mode occurs. In contrast, when $\lambda_0 = 1650$ nm, $\lambda_{guide}$ is 764 nm, which is longer, and thus, the phase-delay mode occurs. The effective wavelength $\lambda_{eff}$ can be obtained from the interval between the successive nodes of the radiation wave. The calculated effective wavelengths for different wavelengths of the incident light are shown in Fig. 2(b). The effective wavelength $\lambda_{eff}$ increases when the wavelength approaches 1550 nm because of the smaller difference between $\lambda_{guide}$ and the structure period. The effective wavelength $\lambda_{eff}$ is near infinity when the incident light wavelength is ≈ 1550 nm, as shown in Fig. 2(b). The effective index $\lambda_{eff}$ can be calculated from the incident light wavelength $\lambda_0$ and effective wavelength $\lambda_{eff}$ as follows:

$$n_{eff} = \frac{\lambda_o}{\lambda_{eff}} \tag{2}$$

The calculated effective index $n_{eff}$ for different incident light wavelengths $\lambda_0$ is shown in Fig. 2(c). At $\lambda_0 \approx 1550$ nm, $n_{eff} \approx 0$ because of the near-infinite effective wavelength $\lambda_{eff}$.

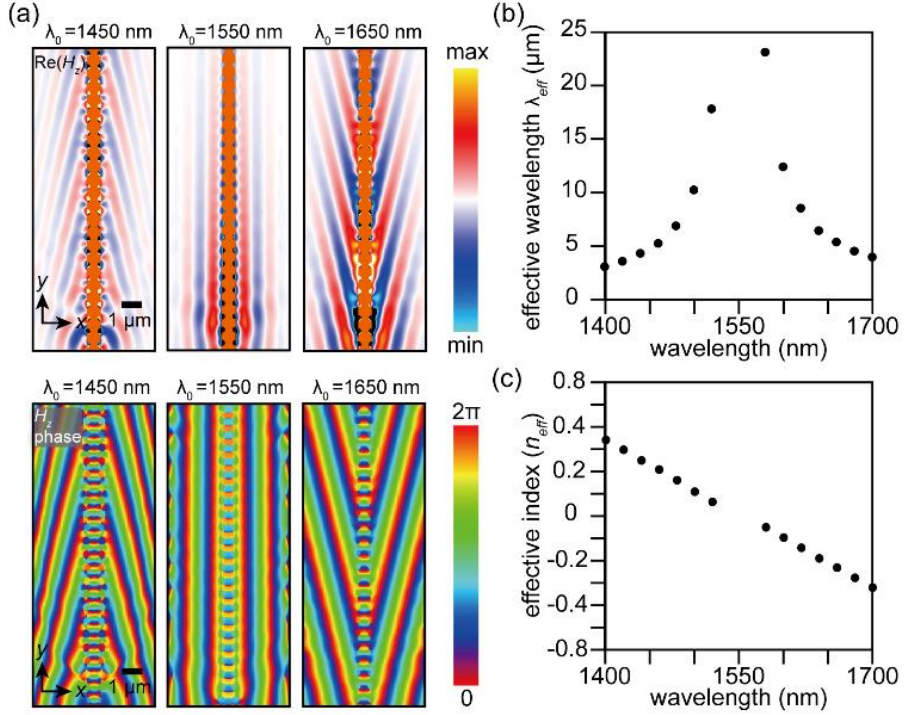

**Fig. 2.** (a) Real part and phase of the magnetic field $H_z$ in the $xy$-plane at the middle of the NZI waveguide at different wavelengths. (b) Calculated effective wavelength $\lambda_{eff}$ and (c) effective refractive index $n_{eff}$ of the NZI waveguide as functions of the incident wavelength $\lambda_0$.

To further investigate the realization of NZI behavior using periodic air-hole structure waveguides, the parameters (period, radius, and thickness) of the NZI waveguide were examined, as shown in Fig. 3. The guide wavelength $\lambda_{guide}$ is determined by the volume-filling fraction of Si, which depends on the air-hole radius for a given period and the incident light wavelength. The effective index $\lambda_{eff}$ and the propagation loss for the air-hole radii = 125, 150, and 175 nm with different incident wavelengths for periods = 640, 660, and 680 nm are shown in Fig. 3 (a), (b), and (c), respectively. The larger the radius, the longer the guide wavelength $\lambda_{guide}$ owing to the smaller volume-filling fraction of Si for a given wavelength and period. Thus, the zero-index wavelength $\lambda_{eff} = 0$ shifted to a shorter wavelength.

Because there was no absorption loss from the material (Si) in the wavelength range of 1400–1700 nm, the propagation loss originated from the light coupling to the free-space through the air holes as well as the bandgap of the photonic crystal-like periodic air-hole structure. The peak of the propagation loss curve corresponds to the high-efficiency light coupling to the free space around the zero-index mode. Notably, contrary to the Dirac-like cone mode approach, which allows only a specific radius/period ratio, the proposed approach allowed different radius/period ratios to enable the realization of a zero index for a target wavelength. For example, a zero index was realized at a wavelength of ≈ 1550 nm when the NZI waveguides with combinations of (period, radius) = (640 nm, 125 nm), (660 nm 150 nm), and (680 nm, 175 nm) were used. Fig. 3(d) shows the zero-index wavelength $\lambda_{eff} = 0$ for different waveguide thicknesses with a period of 660 nm and radius of 150 nm. The thicker the waveguide, the longer the $\lambda_{neff} = 0$ for a given period and radius owing to the larger volume-filling fraction of Si.

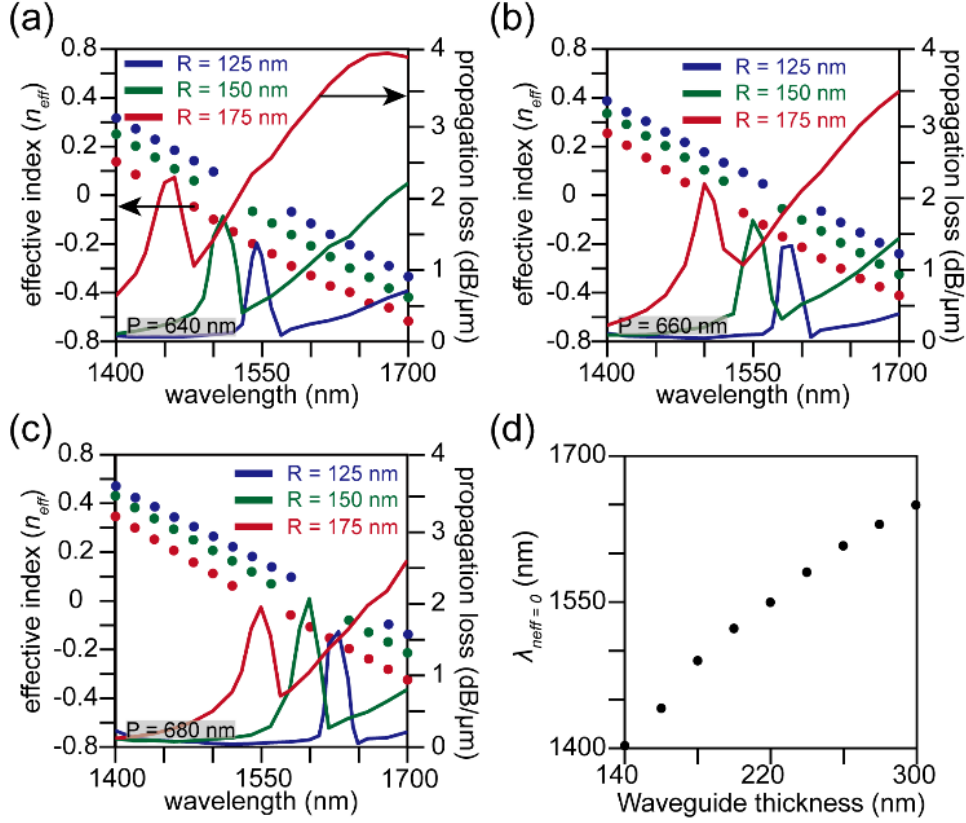

**Fig. 3.** Calculated effective index and propagation loss as functions of the NZI waveguides for different radii with (a) period = 640 nm, (b) period = 660 nm, and (c) period = 680 nm. (d) Zero-index wavelength $\lambda_{neff} = 0$ as a function of the waveguide thickness.

## 4 Beam steering based on near-zero-index waveguide

Owing to the easy coupling of light into the free space through the air holes near the zero-index mode, beam steering can be realized using the NZI waveguide, as shown in Fig. 4. Beam steering in the near field with different incident light wavelengths is illustrated in Fig. 4(a). The radiation in the forward, normal, and backward direction can be realized by tuning the wavelength of the incident light. Fig. 4(b) shows the distribution of the real part of the electric field $E_x$ in the $yz$-plane at the middle of the NZI waveguide at wavelengths of 1450, 1550, and 1650 nm. The dimensions of the NZI waveguide were: period = 660 nm, radius = 150 nm, and thickness = 220 nm. When the incident light has a wavelength of 1550 nm, which is the zero-index wavelength $\lambda_{neff} = 0$, radiation in the normal direction can be observed as shown in Fig. 4(b). The light left perpendicular to the surface of the waveguide indicates that the waveguide is a ZIM at this wavelength. Conversely, the radiation in the forward and backward direction when the incident wavelengths are 1450 and 1650 nm, respectively.

The forward and backward propagating waves correspond to the positive and negative effective refractive indices, respectively, as shown in Fig. 2(c). To obtain a better understanding of the beam-steering properties, the far-field radiation patterns in the *yz*-plane were calculated, as shown in Fig. 4(c). When the light radiates toward the far field, a fan-shaped beam is formed. Although plane waves can be well formed in the near field as forward, normal, and backward radiations, as shown in Fig. 4(b), the inevitable light scattered out from the waveguide contains a fraction of the positive wave vector k along the *y*-direction (as the propagation direction), irrespective of the wavelength of the incident light. Therefore, a fraction of the radiation pattern in the positive *y*-direction, owing to light scattering, would always exist irrespective of the wavelength of the incident light. For the incident light wavelengths of 1450, 1550, and 1650 nm, the beam divergence of the of the beam angle was found to be 28.34°, 38.68°, and 38.23 (full-width at half-maximum, FWHM), respectively. The higher intensity of the radiation in the radiation in the forward direction for an incident light wavelength of 1450 nm is caused by light scattering, as mentioned previously. Owing to the light scattering, forward light scattering can be observed even for the radiation in the normal or backward direction under the incident light wavelengths 1550 or 1650 nm. To evaluate the feasibility of beam steering using NZI waveguides, NZI waveguides made from Ge and $Si_3N_4$ were designed. The Ge and $Si_3N_4$ NZI waveguides were found to exhibit NZI behavior around 1550 nm for the following dimensions: (period, radius, thickness) = (515 nm, 105 nm, 200 nm), and (950 nm, 175 nm, 650 nm), respectively. Fig. 4(d) shows the calculated effective index neff with different incident wavelengths $\lambda_0$ for the NZI waveguides made from Ge, Si, and $Si_3N_4$. Because the wavelength of the guided wave $\lambda_{guide}$ is determined by the refractive index of the material, the wavelength dependence of the effective index $\lambda_{neff}$ ($n_{eff}/\lambda$) varies for different materials. Small and large wavelength dependencies were realized using the low- and high-refractive-index materials $Si_3N_4$ and Ge, respectively. Owing to the weak wavelength dependence of the effective index $n_{eff}$ of the $Si_3N_4$ NZI waveguide, a broadband NZI was realized with a bandwidth of 140 nm. for the central wavelength of 1550 nm. Here, the NZI was defined as $|n_{eff}|<0.1$. In contrast, a narrowband NZI was realized with a bandwidth of 50 nm for a central wavelength of 1550 nm. Fig. 4(e) shows the calculated radiation angle $\theta$ with different wavelengths for the Ge, Si, and $Si_3N_4$ NZI waveguides. Owing to the high refractive index of Ge, a large angular dispersion ($d\theta/d\lambda_0 = 0.26°$) of the radiation was observed. Although the $Si_3N_4$ NZI waveguide exhibited a smaller angular dispersion ($d\theta/d\lambda_0 = 0.01°$) owing to the low refractive index of $Si_3N_4$, a wide steering range of $\Delta\theta \approx 105°$ was achieved. Notably, NZI waveguide-based beam steering was realized with a radiation from forward to backward direction steering angle, which crosses the radiation angle $\theta = 0°$ by changing the wavelength of the incident light.

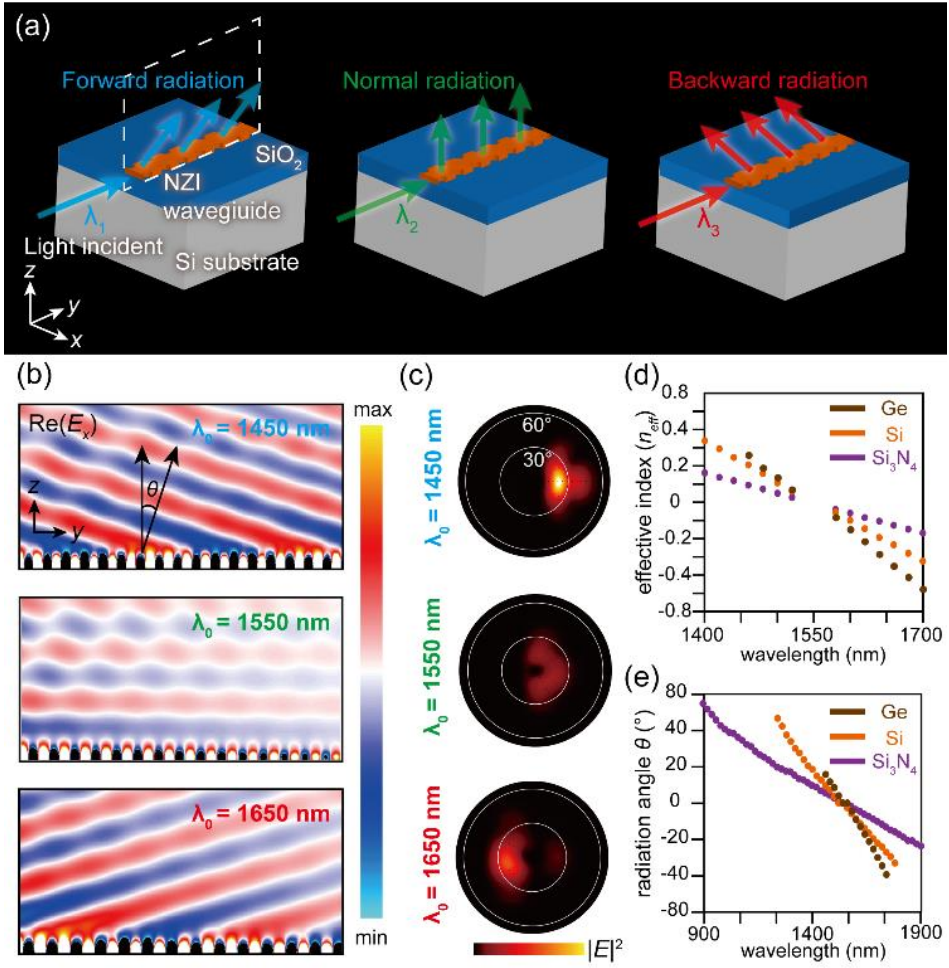

**Fig. 4.** (a) Schematic of beam steering based on NZI waveguides. (b) Real part of the electric field $E_Z$ in the *yz*-plane at the middle of the NZI waveguide at $\lambda_0$ = 1450, 1550, and 1650 nm. (c) Far-field radiation patterns at $\lambda_0$ = 1450, 1550, and 1650 nm. (d) Calculated effective index as a function of $\lambda_0$ for the Ge, Si, and $Si_3N_4$ NZI waveguides. (e) Radiation angle $\theta$ as a function of $\lambda_0$ for the Ge, Si, and $Si_3N_4$ NZI waveguides.

## 5 Conclusion

In summary, a more general and simple design principle that does not require band-structure calculations for the Dirac cone mode was proposed for NZI waveguides. Moreover, owing to the low refractive index of $Si_3N_4$, broadband NZI was achieved with a large bandwidth of 140 nm at telecommunication wavelengths. The proposed design principle of the NZI waveguide facilitates the development of a zero-refractive-index photonic circuit, and beam steering was demonstrated using this NZI waveguide. In contrast to the beam steering based on a slow-light waveguide, the proposed device facilitates a forward-to-backward radiation angle, which crosses the radiation angle $\theta = 0°$, without changing the incident light direction or requiring multiple waveguides. Additionally, a wide steering range ($\theta = -31°–73°$) was achieved using $Si_3N_4$. Such beam-steering functions can be implemented in optical devices, such as security sensors, laser displays, and light detection and ranging systems.

Biographies of the authors are not available.

**Caption List**

**Fig. 1** (a) Schematic of a light source that emits light in a ZIM. (b) Schematic of an NZI waveguide. (c) Schematic of the light propagating in the NZI waveguide and coupling to the free-space light at resonant wavelength. (d) Calculated guide wavelength $\lambda_{guide}$ as a function of the incident light wavelength $\lambda_0$. (e) Distribution of the real part of the magnetic field $H_z$ in the *xy*-plane at the middle of the NZI waveguide.

**Fig. 2** (a) Real part and phase of the magnetic field $H_z$ in the *xy*-plane at the middle of the NZI waveguide at different wavelengths. (b) Calculated effective wavelength $\lambda_{eff}$ and (c) effective refractive index $n_{eff}$ of the NZI waveguide as functions of the incident wavelength $\lambda_0$.

**Fig. 3.** Calculated effective index and propagation loss as functions of the NZI waveguides for different radii with (a) period = 640 nm, (b) period = 660 nm, and (c) period = 680 nm. (d) Zero-index wavelength $\lambda_{neff}=0$ as a function of the waveguide thickness.

**Fig. 4.** (a) Schematic of beam steering based on NZI waveguides. (b) Real part of the electric field $E_Z$ in the *yz*-plane at the middle of the NZI waveguide at $\lambda_0 = $ 1450, 1550, and 1650 nm. (c) Far-field radiation patterns at $\lambda_0 = $ 1450, 1550, and 1650 nm. (d) Calculated effective index as a function of $\lambda_0$ for the Ge, Si, and $Si_3N_4$ NZI waveguides. (e) Radiation angle $\theta$ as a function of $\lambda_0$ for the Ge, Si, and $Si_3N_4$ NZI waveguides.